\documentstyle[here,epsfig]{mn}
\begin{document}
\def\simlt{\mathrel{\rlap{\lower 3pt\hbox{$\sim$}}\raise 2.0pt\hbox{$<$}}}
\def\simgt{\mathrel{\rlap{\lower 3pt\hbox{$\sim$}} \raise 2.0pt\hbox{$>$}}}
\def\di{\mbox{d}}
\def\Msun{M_{\odot}}
\def\HI{\hbox{H$\scriptstyle\rm I\ $}}

\newcommand{\q}{\begin{equation}}
\newcommand{\qa}{\begin{eqnarray}}
\newcommand{\qs}{\begin{eqnarray*}}
\newcommand{\nq}{\end{equation}}
\newcommand{\nqa}{\end{eqnarray}}
\newcommand{\nqs}{\end{eqnarray*}}

\title[The Imprint of the Cosmic Dark Ages on the Near Infrared Background] 
{The Imprint of the Cosmic Dark Ages on the Near Infrared Background}
\author[Ruben Salvaterra \& Andrea Ferrara] 
{R. Salvaterra $^{1}$ \&  A. Ferrara $^{2}$ \\
$^1$SISSA, Via Beirut 4, 34100, Trieste, Italy\\
$^2$Osservatorio Astrofisico di Arcetri, L.go E. Fermi 5, 50125 Firenze, Italy }

\maketitle \vspace {7cm }

\begin{abstract}
The redshifted light of the first (Pop III) stars might substantially 
contribute to the near infrared background (NIRB). By fitting recent data 
with models including up-to-date Pop III stellar spectra, we find that such stars 
can indeed account for the whole NIRB residual  
(i.e. after `normal' galaxy contribution subtraction) if the high redshift 
star formation efficiency is $f_\star=10\%-50\%$, depending on the IMF
(the top-heaviest requiring lowest efficiency) and on the unknown galaxy
contribution in the L band (our models, however, suggest it to be negligible). 
Such epoch of Pop III
star formation ends in all models by $z_{end}\approx 8.8$, with a hard  limit 
$z_{end} < 9$ set by J band observations.
To prevent an associated IGM over-enrichment with heavy elements compared to  
observed levels in the IGM, pair-instability supernovae must be 
the dominant heavy element sources. Alternative explanations must break 
the light-metal production link by advocating very massive stars ($M > 260
M_\odot$) locking their nucleosynthetic products in the compact remnant
or by postulating an extremely inhomogeneous metal enrichment of the Ly$\alpha$
forest. We discuss these possibilities in detail along with the uncertainties 
related to the adopted zodiacal light model.

\end{abstract}

\section{Introduction}

The cosmic infrared background (CIRB) can set important constraints on the
star formation history of the universe (e.g. Hauser \& Dwek 2001 and references
therein). The CIRB consists of the cumulative radiation from all
extragalactic sources. In the near-infrared (NIR) the CIRB is dominated by
starlight, whereas in the mid- and far-infrared it results from dust emission.

Different authors had provided measures of the Near Infrared Background (NIRB)
in the J, K, and L bands using the data of DIRBE instrument aboard of COBE 
satellite.  Recently, Matsumoto et al. (2000) have accurately measured the NIRB 
from 1.4 to 4 $\mu$m, using the data taken with the Near Infrared Spectrometer.
The largest uncertainty on the NIRB measurements is due to the 
subtraction of the interplanetary dust (IPD) scattered sunlight. In spite of
this difficulty some studies have concluded that normal galaxies cannot account  
for the whole observed NIRB (Totani et al. 2001).


Bond, Carr \& Hogan (1986) pointed out that the first population of 
zero-metallicity stars, the so called Population III (Pop III), might
contribute to the cosmic background in the near infrared. More recently,
Santos, Bromm \& Kamionkowski (2001) suggested that a significant part of the unaccounted
NIRB could come from Pop III stars. They considered a very top-heavy 
Initial Mass Function (IMF), assuming that all stars have masses $\simgt 300\;\Msun$ 
as recent three-dimensional numerical simulations seem to suggest 
(Bromm, Coppi \& Larson 1999, 2002; Abel, Bryan \& Norman 2000).  
In this paper we improve on these earlier studies in many respects. First, we
confront theoretical predictions with the recent Matsumoto et al. (2000) 
data, which provides a much tighter test and constraint. Secondly, 
we consider the effect of several IMFs. 
Third, we use the new stellar 
spectra of Schaerer (2002) for zero age mean sequence (ZAMS) metal free stars, 
including the
nebular continuous emission found to be very important for stars with strong 
ionizing fluxes.

We fit the estimate of the unaccounted NIRB finding limits on the 
star formation efficiency and on 
the redshift at which the formation of Pop III stars ends.
We compare our results with high redshift observations and draw some 
results on IGM metal enrichment due to Pop III supernovae.

The paper is organized as follows: in \S 2 we describe the background intensity
calculation, the stellar spectra used, and how we model the IGM. 
In \S 3 the available
data in the NIR range are reviewed along with the number counts of  
normal galaxies from deep field surveys. We discuss the results of our 
analysis in \S 4; in \S 5 we present our conclusions.

We adopt the `concordance' model values for the cosmological parameters: $h=0.7$, 
$\Omega_M=0.3$, $\Omega_{\Lambda}=0.7$, $\Omega_b=0.038$, $\sigma_8=0.9$,
and $\Gamma=0.21$. Here $h$ is the dimensionless Hubble constant, 
$H_0=100h$ km s$^{-1}$ Mpc$^{-1}$; $\Omega_M$, $\Omega_{\Lambda}$, and $\Omega_b$ are 
the total matter, cosmological constant, and baryon density in units of the
critical density; $\sigma_8$ gives the normalization of the power spectrum 
on $8h^{-1}$ Mpc scale and $\Gamma$ is the shape of the power spectrum.
We use the power spectrum for the fluctuations derived by 
Efstathiou, Bond \& White (1992).

\section{COSMIC INFRARED BACKGROUND}

The mean specific intensity of the background  $I(\nu_{0},z_{0})$, as 
seen at frequency $\nu_{0}$ by an observer at redshift $z_{0}$, is 
given by

\q\label{eq:j}
I(\nu_{0},z_{0})= \frac{1}{4\pi}\int^{\infty}_{z_{0}} 
\epsilon(\nu,z)e^{-\tau_{eff}(\nu_{0},z_{0},z)}\frac{\di l}{\di z}\di z 
\nq

\noindent
\cite{Pe}. Here $\epsilon(\nu,z)$ is the comoving specific emissivity, 
$\nu=\nu_0(1+z)/(1+z_0)$, $\tau_{eff}(\nu_{0},z_{0},z)$ is the 
effective optical depth at $\nu_0$ of the IGM between redshift $z_0$ and
$z$, and $\di l/\di z$ is the proper line element

\qs
\frac{\di l}{\di z}=c{[H_0(1+z)E(z)]}^{-1},
\nqs

\noindent
where $c$ is the light speed and

\q\label{eq:e}
E(z)={[\Omega_M(1+z)^3+\Omega_{\Lambda}+
(1-\Omega_M - \Omega_{\Lambda})(1+z)^2]}^{1/2}.
\nq

\subsection{The Comoving Specific Emissivity}

The gas initially virialized in the potential well of the parent
dark matter halo, can subsequently fragment and ignite star formation only if
the gas can cool efficiently and lose pressure support. For a plasma of
primordial composition at temperature $T<10^4$ K, the typical virial temperatures
of the early bound structures, the only efficient coolant is molecular hydrogen.
Thus, a minimum H$_2$ fraction is required for a gas cloud to be able to 
cool in a Hubble time. As the intergalactic relic H$_2$ abundance falls short
by at least two orders of magnitude such requirement, the fate
of a virialized lump depends crucially on its ability to rapidly increase
its H$_2$ content during the collapse phase. Once a critical H$_2$ fractional
abundance of $\sim 5\times10^{-4}$ is achieved in an object, the lump will
cool, fragment and eventually form stars. This criterion is met only by 
larger haloes, so that for each virialization redshift there will exist
a critical mass, $M_{min}(z)$, for which protogalaxies with total mass 
$M_{h}>M_{min}$ form stars and those with mass $M_{h}<M_{min}$ do not. 
We adopt here $M_{min}(z)$ as computed by Fuller \& Couchman (2000).

In absence of additional effects that could prevent or delay the collapse, we
can associate to each dark matter halo with mass $M_{h}>M_{min}$ a 
corresponding stellar mass 

\qa
M_{\star}=f_{\star} \frac{\Omega_b}{\Omega_M} M_{h},
\nqa

\noindent
where $f_{\star}$ is the fraction of baryons able to cool and form stars;
we will refer to this quantity as the star formation efficiency.
Thus, the stellar mass per unit comoving volume at redshift $z$ contained in haloes with
mass $M_h>M_{min}(z)$ is given by

\q\label{eq:rhostar}
\rho_{\star}(z)=\int^{\infty}_{M_{min}(z)}  n(M_{h},z) M_{\star} \di M_{h},
\nq

\noindent
where $n(M_{h},z)$ is the comoving number density of dark matter haloes of mass $M_{h}$ at redshift $z$ given by Press \& Schechter (1974).

\medskip

The comoving specific emissivity, in units of erg s$^{-1}$ Hz$^{-1}$
cm$^{-3}$, is then

\q\label{eq:em}
\epsilon(\nu,z)=l_{\nu}(z) \rho_{\star}(z),
\nq

\noindent
where $\l_{\nu}(z)$ is the specific luminosity of the population 
[in erg s$^{-1}$ Hz$^{-1}$ $\Msun^{-1}$] at redshift $z$ (see Sec. \ref{sec:pop}).

\subsection{The Intergalactic Medium}\label{sec:IGM}

Absorption due to intergalactic gas located in discrete systems along the line of sight 
can seriously distort our view of objects at cosmological distances. 
At wavelengths shortward of Ly$\alpha$ ($\lambda<1216$ \AA) in the emitter
rest frame, the source continuum intensity is attenuated by the combined
blanketing of lines in the Lyman series, and strongly suppressed by the
continuum absorption from neutral hydrogen shortward of the Lyman limit in the
emitter rest frame. As pointed out by Haiman \& Loeb (1999) no flux is transmitted at wavelengths
$\lambda_{0}<\lambda_{\alpha}(1+z_{s})$ from sources at redshift
$1+z_s>32/27(1+z_{i})$, where $z_{i}=6.2$
 (Gnedin 2001) is the currently favored reionization redshift.

The effective optical depth $\tau_{eff}$ through the IGM is defined as 
$e^{-\tau_{eff}}=<e^{-\tau}>$, where the mean is taken over all the lines of
sight to the redshift of interest. For a Poisson distribution of absorbers
(Madau 1991, 1992),

\q
\tau_{eff}(\nu_{0},z_{0},z)= \int^z_{z_0}dz^{\prime}\int^{\infty}_0
dN_{\HI}\zeta(N_{\HI},z^{\prime})(1-e^{-\tau}),
\nq

\noindent
where $\zeta(N_{\HI},z^{\prime})=d^2N/dN_{\HI}dz^{\prime}$ is the distribution 
of the absorbers as a function of redshift and neutral hydrogen column density,
$N_{\HI}$; $\tau(\nu^{\prime})$ is the optical depth of an 
individual cloud for ionizing radiation at frequency $\nu^{\prime}$.

For the Lyman-$\alpha$ forest, in the wavelength range 
$\lambda_{\beta}<\lambda_{0}/(1+z_{em})<\lambda_{\alpha}$ where 
$\lambda_{\alpha}=1216$ \AA$\,$ and $\lambda_{\beta}=1026$ \AA, we have \cite{Ma95}

\qs 
\tau_{eff}=0.0036 \left(\frac{\lambda_{0}}{\lambda_{\alpha}}\right)^{3.46}.
\nqs

When $\lambda_{0}/(1+z_{em})<\lambda_{\beta}$, a significant contribution to
the blanketing opacity comes from the higher order lines of the Lyman series.
In the wavelength range $\lambda_{i+1}<\lambda_{0}/(1+z)<\lambda_i$, the 
total effective line-blanketing optical depth can be written as the sum
of the contributions from the $j\longrightarrow 1$ transitions,

\q
\tau_{eff}=\sum_{j=2,i} A_j \left(\frac{\lambda_{0}}{\lambda_j}\right)^{3.46},
\label{teff}
\nq

\noindent
where $\lambda_j$ and the corresponding values for $A_j$ are given in Table 
\ref{tab_a}.

\begin{table}
\begin{center}
\begin{tabular}{rrr}
\hline
\hline
\multicolumn{1}{c}{j} & \multicolumn{1}{c}{$\lambda_j$ [\AA]} & \multicolumn{1}{c}{$A_j$} \\
\hline
2 & 1216 & 3.6$\times10^{-3}$ \\
3 & 1026 & 1.7$\times10^{-3}$ \\
4 & 973 & 1.2$\times10^{-3}$ \\
5 & 950 & 9.3$\times10^{-4}$ \\
\hline
\end{tabular}
\end{center}
\caption{Coefficients $A_j$ corresponding to $\lambda_j$ in eq. \ref{teff}}
\label{tab_a}
\end{table}

\bigskip

Continuum absorption from neutral hydrogen along the line of sight affects 
photons observed at $\lambda_{0}/(1+z_{em})<\lambda_L$, where $\lambda_L=912$
\AA$\,$  is the Lyman limit. We have

\q
\tau=N_{\HI}\sigma
\nq 

\noindent
where $\sigma(\lambda_{0},z)\sim 6.3\times 10^{-18} (\lambda_{0}/\lambda_L)^3 (1+z)^{-3}$ cm$^2$ (Osterbrock 1989)
is the hydrogen photoionization cross section, and 
$(1+z_c)=(\lambda_{0}/\lambda_L) $
for $\lambda_{0}>\lambda_L$, whereas for $\lambda_{0}<\lambda_L$, \HI 
absorbs photons all the way down to $z_c=0$.

For the redshift and column density distribution of absorption lines, the
usual form can be adopted

\q
\zeta(N_{\HI},z)=\left(\frac{A}{10^{17}}\right)\left(\frac{N_{\HI}}{10^{17} \mbox{ cm$^{-2}$ }}\right)^{-\beta} (1+z)^{\gamma}.
\label{csinhi}
\nq

\noindent
where the values of coefficients  $A$, $\beta$ and $\gamma$ in different 
ranges in $N_{\HI}$ are given in Table \ref{tab_fardal}, and are taken from
Fardal, Giroux \& Shull (1998).

\begin{table}
\begin{center}
\begin{tabular}{llll}
\hline
\hline
\multicolumn{1}{c}{$N_{\HI}$} & \multicolumn{1}{c}{$A$} & \multicolumn{1}{c}{$\beta$} & \multicolumn{1}{c}{$\gamma$} \\
\hline
\multicolumn{1}{c}{$< 10^{14}$} & $1.45\times10^{-1}$ & 1.40 & 2.58 \\
$10^{14}-10^{16}$ & $6.04\times10^{-3}$ & 1.86 & 2.58 \\
$10^{16}-10^{19}$ & $2.58\times10^{-2}$ & 1.23 & 2.58 \\
$10^{19}-10^{22}$ & $8.42\times10^{-2}$ & 1.16 & 1.30 \\
\hline
\end{tabular}
\end{center}
\caption{Best fit value (Model A1) for $A$, $\beta$ and $\gamma$ (Fardal et al. 1998)
used in eq. \ref{csinhi}}
\label{tab_fardal}
\end{table}

\subsection{Scattering of Ly$\alpha$ Photons}\label{sec:lya}

Ly$\alpha$ line photons from first galaxies are absorbed by neutral hydrogen 
along the line of sight (Gunn \& Peterson 1965). However, they are not 
destroyed but scatter and diffuse in frequency to the red of the Ly$\alpha$
resonance owing to the Hubble expansion of the surrounding \HI. Eventually,
when their net frequency shift is sufficiently large, they escape and 
travel freely toward the observer. The profile results in a strong,
asymmetric Ly$\alpha$ emission line around 1225 \AA$\,$ with a scattering tail
extending to long wavelengths.

The resulting scattered line profile, $\Phi(\nu,z)$, has been simulated by Loeb
\& Rybicki (1999). Here, we use the fitting analytical expression given by
Santos et al. (2001)

\q\label{eq:phi}
\Phi(\nu,z)=\left\{
\begin{array}{ll}
\nu_{\star}(z)\,\nu^{-2}\exp\left[\frac{-\nu_{\star}}{\nu}\right] & \mbox{if } \nu>0 \\
0 & \mbox{if } \nu \leq 0
\end{array}
\right. ,
\nq

\q\label{eq:nustar}
\nu_{\star}(z)=1.5\times10^{12} \mbox{Hz} \left(\frac{\Omega_b h^2}{0.019}\right) \left( \frac{h}{0.7} \right)^{-1} \frac{(1+z)^3}{E(z)},
\nq 

\noindent
where $E(z)$ is defined in eq. (\ref{eq:e}). Note that we have allowed 
the IGM in the vicinity of the emitting galaxy to be overdense by a factor
$\delta \approx 10$, to account for the high clustering of first sources.


\subsection{Emission from a Pop III `Stellar Cluster'}\label{sec:pop}

We calculate the emission of a Pop III `stellar cluster' according to the 
equation:

\q
l_{\nu}(z)=\int^{M_{u}}_{M_{l}} F(\nu,M,z) \phi(M)\di M,
\nq

\noindent
where $\phi(M)$ is the IMF normalized so that 
$\int^{M_{u}}_{M_{l}} \phi(M) \di M=1$; $M_{l}$ and $M_{u}$ are
the lower and upper mass limits.
The spectrum of a star of mass $M$ at the rest-frame frequency $\nu$ and
redshift $z$ is given by

\q 
F(\nu,M,z)=l_{\nu}^{star}(M)+l_{\nu}^{neb}(M)+l_{\nu}^{Ly\alpha}(M,z).
\nq

\noindent
where $l_{\nu}^{star}(M)$ is the spectrum of a Pop III star of mass $M$,
$l_{\nu}^{neb}$ is the emission of the nebula surrounding the star, and
$l_{\nu}^{Ly\alpha}(z)$ is the emission due to Ly$\alpha$ line photons 
from a source at redshift $z$ scattered by the IGM (see Sec. \ref{sec:lya}).

As spectra of Pop III we adopt here those modeled by Schaerer (2002)
who relaxed some assumptions made in previous works (Tumlinson \& Shull 2000;
Tumlinson, Giroux \& Shull 2001; Bromm, Kudritzki, \& Loeb 2001). 
Schaerer (2002) presented realistic models for massive Pop III stars
and stellar populations based on non-LTE 
atmospheres, recent stellar evolution tracks and up-to-date evolutionary 
synthesis models. 
Moreover, he included nebular continuous emission, 
which cannot be neglected for
metal-poor objects with strong ionizing fluxes. This process increases 
significantly the total continuum flux at wavelengths redward of Lyman-$\alpha$
and leads in turn to reduced emission line equivalent widths.  

We add thus the contribution of the nebular emission 
($n_e=100$ cm$^{-3}$ is assumed) as described in Schaerer
(2002):

\q
l_{\nu}^{neb}=\frac{\gamma_{tot}}{\alpha_B}(1-f_{esc})q(H)
\nq

\noindent
(e.g. Osterbrock 1989), where $\alpha_B$ (in units of cm$^{-3}$ s$^{-1}$) is the Case B recombination 
coefficient for hydrogen and $f_{esc}$ is the ionizing photon escape fraction out
of the idealized region considered here (we assume $f_{esc}=0$);
$q(H)=Q(H)/M$, where $Q(H)$ is the ionizing photon rate (in units of 
s$^{-1}$) for H and is given for different stellar masses in Table 3 of Schaerer
(2002). The continuous emission coefficient $\gamma_{tot}$, including free-free
and free-bound emission by H, neutral He and single ionized He, as well
as the two-photon continuum of hydrogen is given by
\q
\gamma_{tot}=\gamma_{HI}+\gamma_{2q}+\gamma_{HeI}\frac{n(He^+)}{n(H^+)}+
\gamma_{HeII}\frac{n(He^{++})}{n(H^+)}.
\nq

\begin{figure}
\center{{
\epsfig{figure=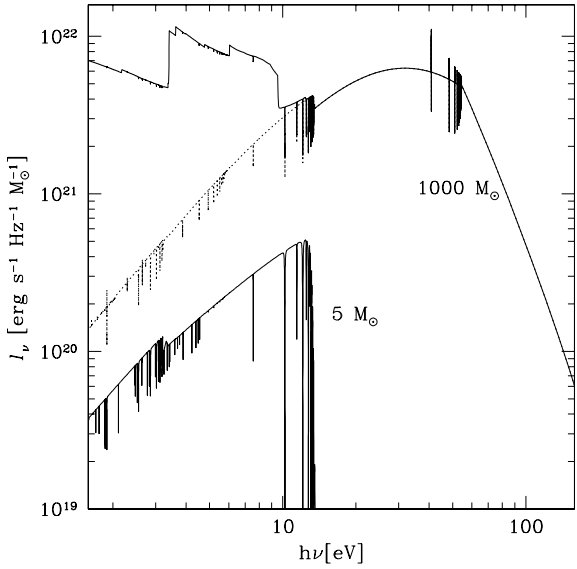,height=8cm}
}}
\caption{\label{fig:spectra_star} Spectra of individual Pop III stars plotted 
as luminosity per unit of stellar mass (in erg s$^{-1}$ Hz$^{-1}$ $\Msun^{-1}$) vs. energy 
(in eV). The spectrum of a 1000 $\Msun$ star with ({\it solid line}) and 
without ({\it dashed}) nebular
emission and of a 5 $\Msun$ are shown. For the 5 $\Msun$ star the
contribution of the nebular emission is negligible.} 
\end{figure}

The continuous emission coefficients $\gamma_i$ (in units of erg cm$^3$ 
s$^{-1}$ Hz$^{-1}$) are taken from Aller (1987) for an
electron temperature of 20000 K. 
The spectrum of a 1000 $\Msun$ star (with and without the nebular emission) 
and of a 5 $\Msun$ star are plotted in Fig. \ref{fig:spectra_star}.  

\bigskip

Finally, the emission of Ly$\alpha$ line photons is given by

\q
l_{\nu}^{Ly\alpha}(z)= c_{Ly\alpha} (1-f_{esc})\,q(H)\, 
\Phi(\nu_{Ly\alpha}-\nu,z),
\nq

\noindent
where $c_{Ly\alpha}=1.04\times 10^{-11}$ erg (Schaerer 2002). 
The line profile $\Phi(\nu)$ is given in eq. (\ref{eq:phi}).

\subsubsection{Initial Mass Function}\label{sec:imf}

The issue of the mass distribution of first metal free stars
has been tackled via various hydrodynamical models and other studies 
(e.g. Abel et al. 1998, Bromm et al. 1999, 2001, 2002, Nakamura \& Umemura 2001,
Ripamonti et al. 2002).
There seems to be an overall consensus that massive stars (up to 1000 $\Msun$)
may form. Simulations seem to suggest 
that the primordial IMF might have been biased towards stellar masses $\geq 100$ 
$\Msun$, but other studies (e.g. Nakamura \& Umemura 2001) found that the
formation of stars with masses down to 1 $\Msun$ is not excluded.

In the local universe the IMF is usually well described by the standard 
Salpeter law (Salpeter 1955)

\q
\phi(M)\propto M^{-2.35}.
\nq

Larson (1998) suggested a different form for the high $z$ IMF. This has a universal Salpeter-like form at
the upper end, but flattens below a characteristic mass, $M_{c}$ which
may vary with time. By increasing $M_{c}$ we can mimick 
a top-heavy IMF of Pop III stars,

\q
\phi(M)\propto M^{-1}\left( 1+\frac{M}{M_{c}} \right)^{-1.35}.
\nq

Hernandez \& Ferrara (2001) have explored the predictions of the standard 
hierarchical clustering scenario of galaxy formation, regarding the numbers 
and metallicities of Pop III stars that are likely to be found within our
Galaxy today. By comparing these values with observational data, they 
suggested that the IMF of first stars was increasingly high-mass weighted 
towards high redshifts, levelling off at $z\ge 9$ at a characteristic 
stellar mass scale of 10-15 $\Msun$.

\begin{figure}
\center{{
\epsfig{figure=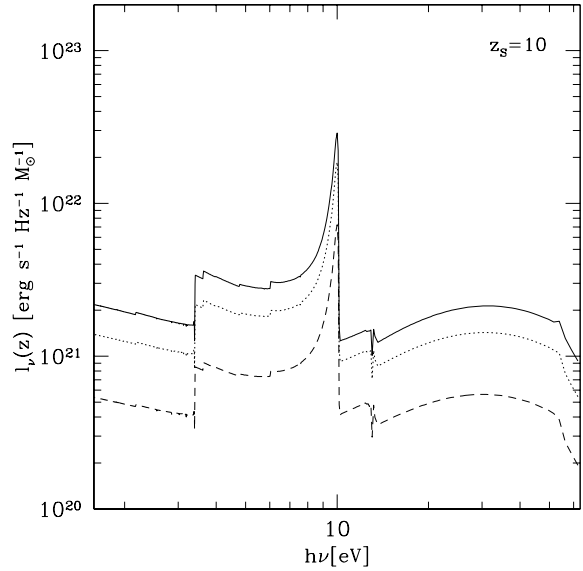,height=8cm}
}}
\caption{\label{fig:pop3} Pop III stellar cluster spectra for different IMFs. 
{\it Dashed line}: Salpeter. {\it Dotted line}: heavy IMF. {\it Solid line}: 
very heavy IMF. The sharp peak around 1220 \AA$\,$ is due to  
Ly$\alpha$ nebular emission for sources at $z_s=10$, filtered
through the IGM. The flux at E$>13.6$ eV is then strongly suppressed by the 
absorption due to the intergalactic gas (see Sec. \ref{sec:IGM})}. 
\end{figure}

\bigskip
In view of our poor knowledge of the IMF at high redshift we consider here 
three different distributions of mass for the first population of stars: 
Salpeter, Larson with $M_{c}=15$ $\Msun$ (heavy IMF), and Larson with 
$M_{c}=100$ $\Msun$
(very heavy IMF). For all IMFs, $M_l=1$ $\Msun$ and $M_u=1000$ $\Msun$. 
In Fig. \ref{fig:pop3} are shown the resulting spectra of a
Pop III cluster at redshift $z_s$=10 for the three different IMFs assumed.

\section{Observational Constraints}\label{sec:data}

\begin{figure}
\center{{
\epsfig{figure=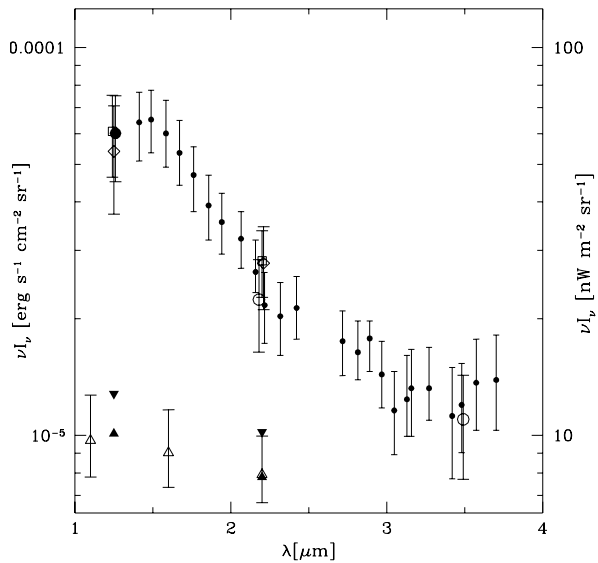,height=8cm}
}}
\caption{\label{fig:nir} Available NIR data. 
The small filled circles are the NIRS data (Matsumoto et al. 2000). The open symbols
are the DIRBE results: squares for Wright 
(2001), diamonds for Cambresy et al. (2001), and circles for Gorijan, Wright 
\& Chary 
(2000). The big filled circle is the Kiso star count measurement.
The data are slightly offset for clarity. The errors are at 1$\sigma$ and for
all the data the Kelsall et al. (1998) model for the zodiacal light is applied.
The open triangles are the count integration from the {\it Hubble Deep Field} 
(Madau \& Pozzetti 2000), whereas the filled triangles report upper and lower 
limits on the count integration from {\it Subaru Deep Field} when also the 
contribution of missed galaxies is considered (Totani et al. 2001).}
\end{figure}

Observations of the NIRB are seriously hampered by the
strong atmospheric foreground (e.g. Mandolesi et al. 1998). 
On the other hand space measurements are 
not easy because it is difficult to subtract correctly the contribution to the
extragalactic background light by interplanetary dust scattered sunlight
(zodiacal light). Using the data from the instrument DIRBE aboard of the
COBE satellite and from the 2MASS data, Cambresy et al. (2001) found a 
non zero, isotropic background in the J and K bands. With the same data but a
different zodiacal light model, Wright \& Johnson (2001) extended the flux
estimate to the L band. Cambresy's and Wright's results in the J and K bands 
are compatible if the same zodiacal light model is applied. 
Matsumoto et al. (2000) provided a preliminary analysis of 
the the NIRS data obtaining a detection of the NIRB in the 1.4-4 
$\mu$m range.
They applied the same zodiacal light model of Cambresy et al. (2001) and 
obtained compatible results in the DIRBE bands.
At 8000 $\AA$, Bernstein, Freedman \& Madore (2002) observed a CIRB of 
$1.76\pm0.48\times10^{-5}$ erg s$^{-1}$ cm$^{-2}$ sr$^{-1}$, that is 
$\sim$4 times smaller than the DIRBE and NIRS 
measurements in the J band, showing a break around $\sim$1 $\mu$m. 
The integrated counts in the J and K bands from deep surveys do not account for
the total NIRB (Madau \& Pozzetti 2000; Totani et al. 2001). Totani 
et al. (2001) modeled the contribution of galaxies missed 
by deep galaxy surveys. They found unlikely that the contribution 
of all the normal galaxies to the NIRB is larger than 30\%. 
All the available data are plotted in Fig. \ref{fig:nir}. In Table 
\ref{tab:nir} the estimate of the NIRB from the 
DIRBE data (top panel) and the contribution of normal galaxies (bottom panel)
are reported.

\begin{table*}
\begin{center}
\begin{tabular}{ccccl}
\hline
\hline
J band  & K band & L band & M & Reference \\
(1.25 $\mu$m) &  (2.2 $\mu$m) & (3.5 $\mu$m) &  & \\
\hline
& (16.2$\pm$6.4) & &  W & Gorijan et al. 2000 \\
& 22.4$\pm$6.0 & 11.0$\pm$3.3 &  K  & \\
\hline
 & (23.1$\pm$5.9) & (12.4$\pm$3.2) & W & Wright \& Reese 2000 \\
\hline
(27.7$\pm$14.5) &  (19.9$\pm$5.3) & &    W & Wright 2001 \\
 60.8$\pm$14.5 & 28.2$\pm$5.5 & & K & \\
\hline
 54.0$\pm$16.8 & 27.8$\pm$6.7 & & K & Cambresy et al. 2001 \\
\hline
(24.3$\pm$18) & (24$\pm$6) & (13.8$\pm$3.4) & W & Wright \& Johnson 2001 \\
\hline
 60.1$\pm$15 & & &  K & Kiso star counts \\
\hline
\hline
\multicolumn{5}{c}{Simple Integration of Observed Galaxy Counts} \\
\hline
 $(9.7^{+3.0}_{-1.9})^a$ & $7.9^{+2.0}_{-1.2}$ & &  & Madau \& Pozzetti 2000 \\
$10.9\pm1.1$ & $8.3\pm0.8$  & & & Totani et al. 2001 \\
\hline
\multicolumn{5}{c}{Estimated Resolved Fraction} \\
\hline
0.97 & 0.93 & & & Model A \\
0.95 & 0.92 & & & Model B \\
\hline
\multicolumn{5}{c}{Contribution to the NIRB of All Normal Galaxy} \\
\hline
 10.1-12.8 & 7.8-10.2 & & &  Totani et al. 2001 \\
\hline
\end{tabular}
\end{center}
\caption{Summary of the recent observations of the NIRB from the DIRBE data (top panel). 
Units are nWm$^{-2}$sr$^{-1}$ (or $10^{-6}$ erg s$^{-1}$ cm$^{2}$ sr$^{-1}$). 
In the fourth column is given the Model used to 
subtract the Zodiacal Light: 'W' stands for Wright (1998) and  'K' is for
Kelsall et al. (1998). In the bottom panel  are shown the results of count 
integration from the {\it Hubble Deep Field} (Madau \& Pozzetti 2000) and from 
the {\it Subaru Deep Field} (Totani et al. 2001) surveys, the estimate of 
the contribution of missed galaxies (Totani et al. 2001. Model A: number 
evolution with $\eta\sim 1$; Model B: no number evolution with $\eta=0$) 
and the total flux from all normal galaxies (Totani et al. 2001).}
\begin{flushleft} 
{\footnotesize $^a$ Estimate at slightly different wavelengths}
\end{flushleft}
\label{tab:nir}
\end{table*}                   

\subsection{The DIRBE Data}

The Diffuse Infrared Background Experiment (DIRBE) on board of the Cosmic
Background Explorer (COBE, see Boggess et al. (1992)) satellite was designed 
to search for the 
Cosmic Infrared Background from 1.25 $\mu$m to 240 $\mu$m. The DIRBE
instrument (Silverberg et al. 1993) was an absolute photometer which 
provided maps of the full sky in 10 broad bands at 1.25, 2.2, 3.5, 4.9, 12,
25, 60, 100, 140, and 240 $\mu$m with a starlight rejection of $< 1$ 
nW m$^{-2}$ sr$^{-1}$ and an absolute brightness calibration uncertainty of 0.05
and 0.03 nW m$^{-2}$ sr$^{-1}$, at 1.25 $\mu$m and 2.2 $\mu$m respectively.
A summary of the DIRBE results is provided by Hauser et al. (1998) and
all the NIRB measures based on the DIRBE data are reported in Table 
\ref{tab:nir} (top panel).

In the near infrared bands the dominant foreground intensities in the DIRBE 
data are the zodiacal light and the light from stars in the Milky Way.

A first attempt to detect the NIRB from the DIRBE data was 
carried out by the DIRBE group but this gave only an upper limit and failed the 
tests of isotropy even in limited regions of sky (Arendt et al. 1998). 
Dwek \& Arendt (1998) made a correlation study for the K and L bands and obtained
a lower limit for the L band. 

Gorjian et al. (2000) removed the foreground due to Galactic stars by directly
measuring all stars brighter than  9th magnitude at 2.2 and 3.5 $\mu$m
in a 2$^{\circ}\times 2^{\circ}$ dark spot near the North Galactic Pole using 
ground-based telescopes. They calculated the contribution of fainter stars 
using the statistical model of Wainscoat et al. (1992) and subtracted the
zodiacal light contribution using a improvement of the 
Wright (1998) model. Gorjian et al. found significant positive residuals
in the K and L bands which they identified as a probable detection of the
NIRB. 
Wright \& Reese (2000) obtained a consistent estimate of the NIRB in the
same bands using a different approach based on a histogram fitting method to
remove the stellar foreground from the DIRBE data.

Wright (2001) used the Two Micron All Sky Survey (2MASS) data (Cutri et al.
2000) to remove the contribution of Galactic stars brighter than 14th 
magnitude from the DIRBE maps at 1.25 and 2.2 $\mu$m in four dark regions
in the north and south Galactic Pole caps. For the subtraction of the zodiacal
light foreground the model presented in Gorjian et al. (2000) is applied.
Cambresy et al. (2000) have also used the 2MASS data to remove the Galactic
stars from the DIRBE data, but they modeled the zodiacal light 
as in Kelsall et al. (1998). Their results agree with those of Wright (2001) 
if the same zodiacal light model is applied.

Recently, Wright \& Johnson (2001) extended the analysis of Wright (2001) to
13 fields with a wide range of ecliptic latitudes tripling the number of 
pixels. They obtained an estimate of the NIRB at 1.25 and 2.2 $\mu$m consistent
with the previous results. They also found a significant residual in the L band
combining the Wright (2001) and Dwek \& Arendt (1998) techniques, consistent
with the Wright \& Reese (2000) result.

Matsumoto et al. (in preparation) observed in the J band with 
the Kiso Schmidt Telescope toward the DIRBE dark spot. They subtracted all
detected stars brighter than 14th magnitude and estimated the contribution
of fainter stars by the Cohen's sky model (Cohen 1997). They used the 
Kelsall's zodiacal light model.

\subsection{The NIRS Data}

The Near Infrared Spectrometer (NIRS)\footnote{http://www.ir.isas.ac.jp/irts/nirs/index-e.html} 
is one of the focal instruments of the InfraRed Telescope in Space (IRTS)
(Noda et al. 1994). The NIRS covers the 
wavelength range from 1.4 to 4.0 $\mu$m with a spectral resolution of 0.13
$\mu$m. The beam size is $8^{\prime}\times 20^{\prime}$. which is considerably
smaller than that of DIRBE; about 7\% of the sky is surveyed. In order to reduce
the contribution from the faint stars, the sky at high latitudes 
($b>40^{\circ}$) is chosen.

Matsumoto et al. (2000) provided a preliminary analysis of the NIRS data,
reporting detection of the Cosmic Infrared Background based upon 
analysis of the five days of data which were least disturbed by atmospheric,
lunar, and nuclear radiation effects. The sky area analyzed included Galactic
latitudes from 40$^{\circ}$ to 58$^{\circ}$, and ecliptic latitudes from 
12$^{\circ}$ to 71$^{\circ}$.
NIRS is able to identify stars brighter than 10.5 mag. at 2.24 $\mu$m, whereas to find out the 
contribution of fainter stars the Cohen model (1997) is used. The model of
Kelsall et al. (1998) is applied to subtract the contribution of
the zodiacal light foreground, interpolating between the DIRBE wavelengths to
the wavelengths of the NIRS measurements.
After subtraction of the IPD contribution, there remains a fairly isotropic
residual emission, which they interpreted as evidence for a non-zero 
background.
To obtain a quantitative value for the background at each wavelength they
correlated their star-subtracted brightness at each point with the IPD model
brightness, and used the extrapolation to zero IPD contribution as a
measurement of the NIRB.

The NIRB intensities reported by Matsumoto et al. (2000) near 2.2 and 3.5 $\mu$m are
similar to the values found by Gorjian et al. (2000). At shorter wavelengths,
the NIRS results continue to rise steeply to $\sim 6.5\times10^{-5}$ 
erg s$^{-1}$ cm$^{-2}$ sr$^{-1}$ at 1.4 
$\mu$m. This is somewhat above the 95\% CL upper limit at 1.25 $\mu$m of 
Wright (2001), but it is in agreement with the value obtained by Cambresy et 
al. (2001) using the same zodiacal light model. The NIRS result implies an 
integrated background energy over the 1.4-4.0 $\mu$m range of 
$\sim3.0\times10^{-5}$ erg s$^{-1}$ cm$^{-2}$ sr$^{-1}$.

The preliminary report by Matsumoto et al. (2000) does not provide details
regarding systematic uncertainties in their results. The largest uncertainty
in the reported NIRB values is attributed to the IPD model.

\subsection{Galaxy Contribution from Deep Counts}

Deep optical and NIR galaxy counts provide an estimate of the extragalactic
background light (EBL) coming from normal galaxies in the universe.
Madau \& Pozzetti (2000) derived the contribution of known galaxies in the
UBVIJHK bands from the {\it Southern Hubble Deep Field}. 
Although the slope of the number-magnitude relation of the faintest counts
is flat enough for the count integration to converge, the recent observations
of the NIRB (see previous Section) suggest that the diffuse EBL
flux is considerably larger than the count integration. However, a considerable
fraction of EBL from galaxies could still have been missed in deep galaxy 
surveys because of various selection effects.

Totani et al. (2001) using the {\it Subaru Deep Field} data (Maihara et al. 2001)
have shown that more than 80\%-90\% of EBL from galaxies has been resolved in
the K and J bands and that the contribution by missing galaxies cannot 
account for the discrepancy between the count integration and the NIRB
observations. In Table \ref{tab:nir} (bottom panel) we show the count 
integration in optical and NIR bands from Madau \& Pozzetti (2000) and Totani
et al. (2001), and the resolved fractions for two galaxy evolution models.
The possible number evolution of galaxies is considered by a phenomenological
model in which the Schechter parameters of the luminosity function have a 
redshift dependence as $\phi^*\propto (1+z)^{\eta}$ and 
$L^*\propto (1+z)^{-\eta}$; i.e., luminosity density is conserved. In
Model A an evolution with $\eta= 1$ is considered, whereas in Model B 
no evolution ($\eta=0$) is assumed (see Totani et al. (2001) for more 
details). Currently no measure of the contribution of normal galaxies to the
NIRB in L band is available whose accurate measurement has to await for SIRTF. 

\section{Model vs. Observations}

We compare the NIR data with the predictions of our model, summarized by eq. 
(\ref{eq:j}). The data set is composed by the measurements of Matsumoto et al. 
(2000) and those obtained by the DIRBE data using the Kelsall model for the 
zodiacal light emission as reported in Table \ref{tab:nir} (top panel).
For each wavelength we subtracted from the data the contribution of 
normal galaxies using the upper (lower)
limits in the J and K band, obtained by Totani et al. (2001) (last line of 
Table \ref{tab:nir}). 
As discussed already, for $\lambda> 2.2$ $\mu$m no estimate of the NIRB due to normal galaxies is 
available. Thus in the range 2.2 to 4 $\mu$m  we consider two extreme cases: 
no contribution from normal galaxies (labeled with {\bf cut}) and 
a constant contribution (labeled with {\bf flat}).

\subsection{Model Results}

\begin{table}
{\small
\begin{center}
\begin{tabular}{lcccc}
\hline
\hline
IMF & galaxy & \multicolumn{3}{c} {$f_\star$} \\
 & model & $z_{end}=8$ & $z_{end}=9$ & $z_{end}=10$ \\
\hline
Salpeter  & cut & 0.37$\pm$0.03 & 0.59$\pm$0.05 & 0.89$\pm$0.08 \\
 & flat & 0.24$\pm$0.03 & 0.39$\pm$0.05 & 0.58$\pm$0.08 \\
\hline
Heavy & cut & 0.15$\pm$0.01 & 0.24$\pm$0.02 & 0.36$\pm$0.03  \\
 & flat & 0.10$\pm$0.01 & 0.15$\pm$0.02 & 0.23$\pm$0.03 \\
\hline
Very & cut & 0.10$\pm$0.01 & 0.15$\pm$0.01 & 0.23$\pm$0.02 \\
~heavy & flat &  0.06$\pm$0.01 & 0.10$\pm$0.01 & 0.15$\pm$0.02 \\
\hline
\end{tabular}
\end{center}
}
\caption{Values of $f_{\star}$ from the fit to the data with different IMFs. 
The quoted statistical errors are at 95\% C.L..}
\label{tab:res}
\end{table}

\begin{figure}
\center{{
\epsfig{figure=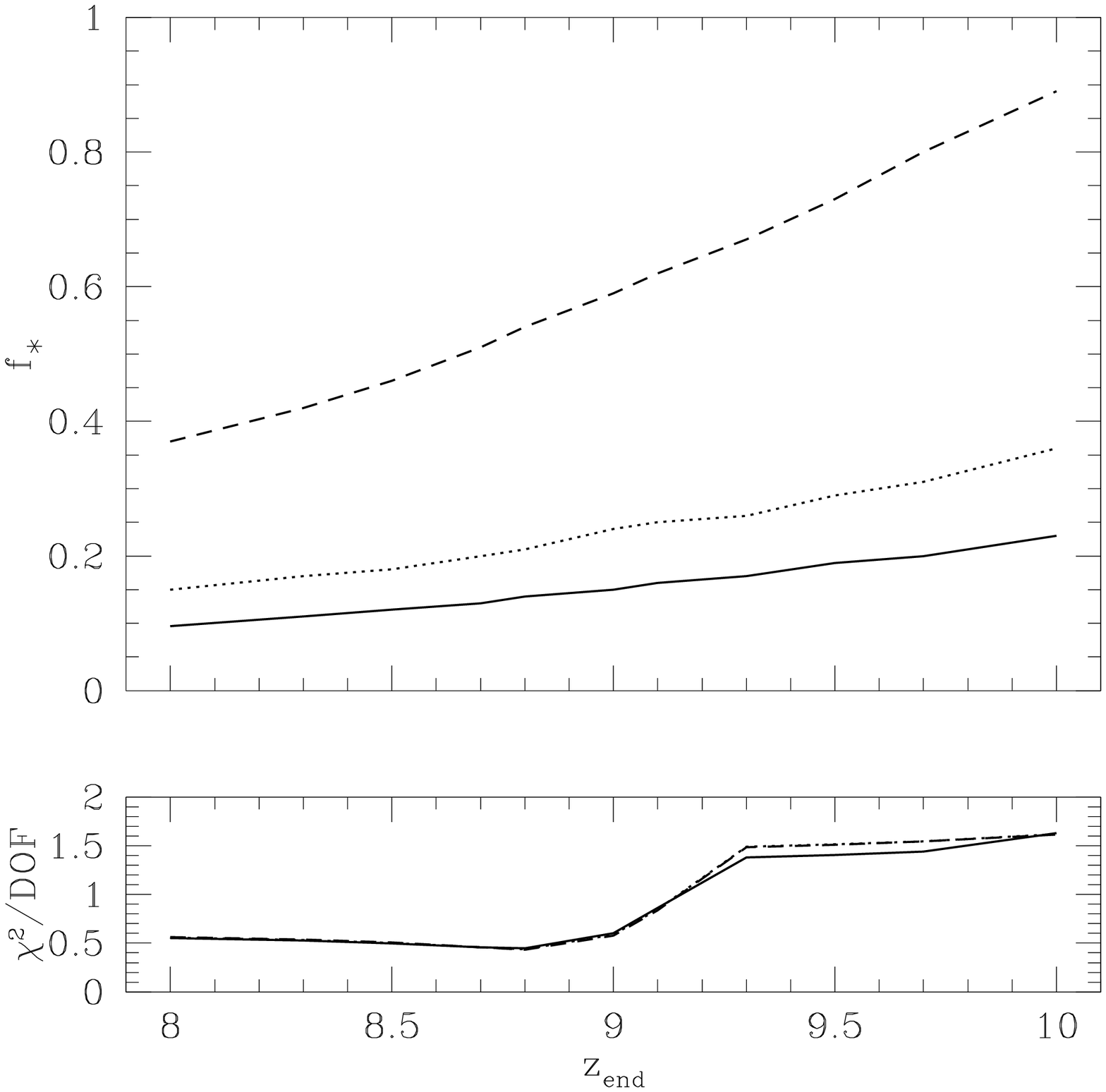,height=8cm}
}}
\caption{\label{fig:fstar} Star formation efficiency $f_\star$ accounting
for the whole unexplained NIRB, as a function of
redshift at which the formation of Pop III stars ends. The model {\bf cut} for the counts integration for $\lambda>2.2$ $\mu$m is adopted.
{\it Dashed line}: Salpeter IMF. 
{\it Dotted line}: heavy IMF.
{\it Solid line}: very heavy IMF. 
The bottom panel shows the value of $\chi^2$/DOF.}
\end{figure}
 
The fit results are quite sensitive to the adopted IMF
and to the way we model the contribution of normal galaxies for 
$\lambda > 2.2$ $\mu$m. In Table \ref{tab:res} are reported the results of the fit for different IMFs, count 
integration model, and $z_{end}$. The reported errors represent  the
statistical uncertainty of the fit, but systematic uncertainties difficult
to evaluate might be present.             
In Fig. \ref{fig:fstar} are plotted the values of the star formation 
efficiency and the corresponding $\chi^2$/DOF (Degree of Freedom)
as function $z_{end}$ for the {\bf cut} galaxy count model.
The $\chi^2$ increases rapidly as $z_{end}$ becomes greater than 9, showing
that in this case our model fails to account of the NIRB observation in the
J band.

For a Salpeter IMF we find that a large fraction of baryons have to be turned 
into stars to account of the whole unexplained NIRB. This fraction goes
from ~37\% ($z_{end}=8$) to ~89\% ($z_{end}=10$) if we consider the case of no 
contribution from normal galaxies for $\lambda>2.2$ $\mu$m, slightly dependent
on the choice of the upper and lower limits of the count integration. 
If we consider a constant ({\bf flat} model) contribution of normal galaxies up to 2.2 $\mu$m 
the value of $f_\star$ goes from $\sim0.24$ ($z_{end}=8$) to $\sim0.58$
($z_{end}=10$).
The value of $f_\star$ decreases considerably if we consider a top-heavy IMF.
For the {\bf cut} model of the normal galaxy contribution, $f_\star$ ranges 
between $\sim 0.15$ ($z_{end}=8$) and $\sim 0.36$ ($z_{end}=10$) if $M_{c}=15$ 
$\Msun$ and between $\sim 0.1$ and $\sim 0.23$ if $M_{c}=100$ $\Msun$, whereas
for the {\bf flat} model we find $f_\star$ values from
$\sim 0.1$ ($z_{end}=8$) to $\sim 0.23$ ($z_{end}=10$) with a heavy IMF, and 
$\sim 0.06-0.15$ (for $z_{end}=8-10$) with a very heavy IMF.

\begin{figure}
\center{{
\epsfig{figure=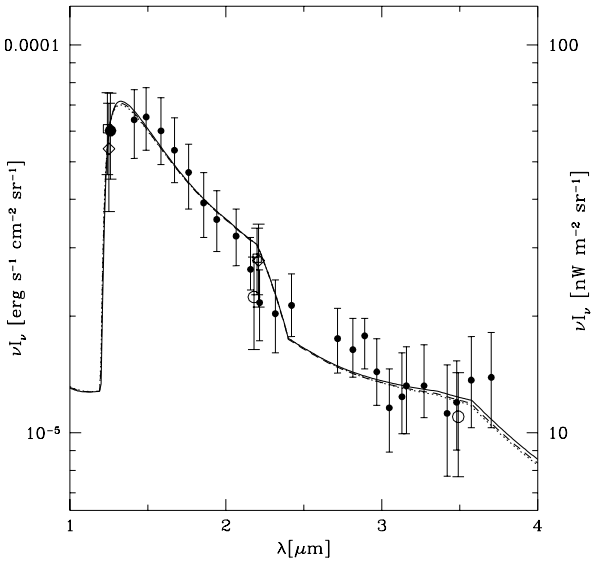,height=8cm}
}}
\caption{\label{fig:fit} Best results of the fit for different IMFs. 
{\it Dashed line}: Salpeter IMF  $z_{end}=8.79$, $f_\star=0.53$. 
{\it Dotted line}: heavy IMF $z_{end}=8.8$, $f_\star=0.21$ .
{\it Solid line}: very heavy IMF $z_{end}=8.82$, $f_\star=0.14$. 
The model {\bf cut} for the counts integration for $\lambda>2.2$ $\mu$m is adopted. The data are explained in Sec. \ref{sec:data} and in Fig. \ref{fig:nir}}
\end{figure}

To find the best model, we allow both $f_\star$ and $z_{end}$
to vary simultaneously and perform a two-parameter fit to the NIR data. 
The results are reported in Table \ref{tab:bf}. 
Again, quoted errors take into account for statistical uncertainties of the fit.
In Fig. \ref{fig:fit}
we plot  the best fit models for the {\bf cut} contribution of normal 
galaxies. For the {\bf flat} model the star formation efficiencies are 
significantly lower than in the {\bf cut} case, but the $\chi^2$/DOF 
remains always greater than 1.9, showing 
that a low contribution of normal galaxies for $\lambda>2.2$ $\mu$m is 
favored by our model.
A very remarkable result is the constancy of $z_{end} \approx 8.8$ in
the various cases,  independent on the IMF or the galaxy contribution model: 
its value is strongly constrained by the position of the break around 1 $\mu$m.
The break is principally due to the Ly$\alpha$ line emission redshifted
into the J band, so that $\lambda_{obs}\sim \lambda_{\mbox{Ly}\alpha}/(1+z_{end})$.


\begin{table}
\begin{center}
\begin{tabular}{lccc}
\hline
\hline
IMF & galaxy & $f_\star$ & $z_{end}$ \\
 & model & & \\ 
\hline
Salpeter  & cut & 0.53$\pm$0.06 & 8.79$\pm$0.10 \\
& flat & 0.32$\pm$0.04 & 8.69$\pm$0.01 \\
\hline
Heavy & cut & 0.21$\pm$0.02 & 8.80$\pm$0.01 \\
 & flat & 0.13$\pm$0.02 & 8.75$\pm$0.01 \\
\hline
Very & cut & 0.14$\pm$0.01 & 8.83$\pm$0.01 \\
~heavy & flat & 0.09$\pm$0.01 & 8.79$\pm$0.01 \\
\hline
\end{tabular}
\end{center}
\caption{Best fit values for $f_{\star}$ and $z_{end}$ with different IMFs. 
The quoted statistical errors are at 95\% C.L..}
\label{tab:bf}
\end{table}

\smallskip

Barkana (2002) derived a value for $f_{\star}$ using the measured 
distribution of star formation rates in galaxies
at various redshifts (Lanzetta et al. 2002) and semi-analytic models of 
hierarchical galaxy formation in a $\Lambda$CDM cosmology. He found a 
best fit value of 2.3$^{+0.8}_{-0.5}$\% (2$\sigma$), neglecting dust 
obscuration, but the uncertainties on this evaluation are still large. 
If this value applies also to the first star formation, 
for a Salpeter IMF the Pop III contribution to the NIRB is not significant
($< 10$\%), whereas Pop III stars characterized by a top-heavy IMF 
continue to contribute substantially to the NIRB.

\subsection{Associated Metal Enrichment}

Recent theoretical analyses on the evolution of metal-free stars predict that
the fate of the massive metal-free stars can be classified as follows (Heger,
Woosley \& Waters 2000, Chiosi 2000, Heger \& Woosley 2002):

\begin{enumerate}
\item $M>260$ $\Msun$: the nuclear energy release from the collapse of stars 
in this mass range is insufficient to reverse the implosion. The final result 
is a very massive black hole (VMBH) locking up all heavy elements produced;
\item 130 $\Msun$ $<M< 260$ $\Msun$: the mass regime of the pair-instability 
supernovae (SN$_{\gamma\gamma}$). Precollapse winds and pulsations should result in little mass loss,
the star implodes reaching a maximum temperature that depends on its mass and then 
explodes, leaving no remnant. The explosion expels metals into the surrounding 
ambient ISM;
\item 40 $\Msun$ $<M< 130$ $\Msun$: black hole formation is the most likely 
outcome, because either a successful outgoing shock fails to occur or the 
shock is so weak that the fall-back converts the neutron star remnant into a 
black hole (Fryer 1999);
\item 8 $\Msun$ $<M< 40$ $\Msun$: SN explosion from `normal' progenitors.
\end{enumerate}
 
Stars in the mass range (i) and (iii) above fail to eject most of (or all) 
their heavy elements, whereas (ii) and (iv) type of stars eject metals into 
the IGM. 

To estimate the metal production of Pop III objects we 
consider SN yields of Woosley \& Weaver (1995) for progenitor metal-free 
stars between 
8 and 40 $\Msun$ and the results of Heger \& Woosley (2002) for very massive 
objects of pair instability SNe originating from stars in the mass range $\sim 130 
M_\odot$ to 260 $\Msun$. We neglect the contribution from long lived intermediate mass 
stars ($1<M<8$ $\Msun$) as their evolution time scale is longer than the Hubble
time at the relevant redshift. The density of metals ejected into the IGM in 
units of the critical density, $\rho_c=3H_0^2/8\pi G$, is obtained by

\q
\Omega_Z(z)=f_Z(\phi)\; \Omega_{\star}(z)
\nq

\noindent
where $f_Z$ is the fraction of metal ejected from normal and pair-instability 
supernovae; it depends on the adopted IMF, $\phi$. Finally, 
$\Omega_{\star}(z)=\rho_{\star}/\rho_c$, where $\rho_{\star}$ is given in 
eq. (\ref{eq:rhostar}).

\medskip

\begin{figure}
\center{{
\epsfig{figure=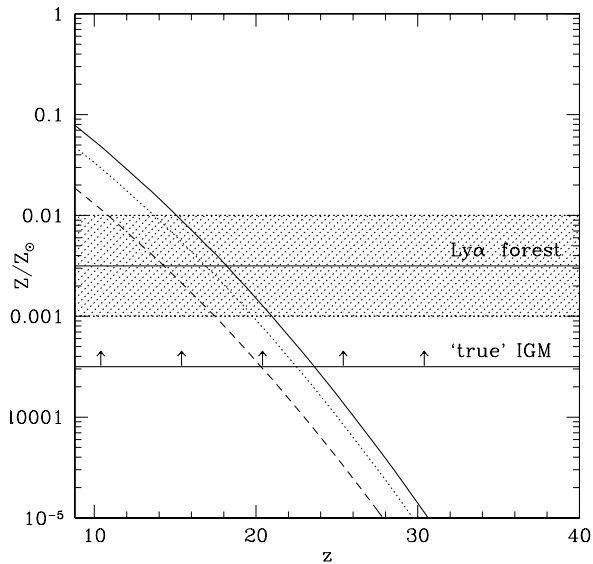,height=8cm}
}}
\caption{\label{fig:metal} IGM metallicity as a function of redshift. {\it Dashed line}: Salpeter IMF. 
{\it Dotted line}: heavy IMF. {\it Solid line}: very heavy IMF. 
The curves are obtained imposing that the whole unaccounted NIRB is
due to Pop III stars and the best results for star formation efficiency of 
Fig. \ref{fig:fit} are assumed. The horizontal solid lines represent the value
of the metallicity in the Ly$\alpha$ forest (Dav\'e et al. 1998) and in the 
`true' IGM (Songaila 2002).}
\end{figure}

We calculate the IGM metallicity by imposing that the whole unaccounted
NIRB is due to Pop III stars. We use the best results for the 
efficiency reported in Fig. \ref{fig:fit} for the {\bf cut} model of the
galaxy counts. We assume here that all the metal escape into the IGM and that
they are smoothly distributed. 
Fig. \ref{fig:metal} shows the metallicity of the IGM (in solar units) as a 
function of redshift. 

We compare our results 
with the estimated value of metals in the Ly$\alpha$ forest 
($10^{14.5}< N_{\HI}<10^{16.5}$ cm$^{-2}$) at $z\sim 3$ (Songaila \& Cowie
1996; Dav\'e et al. 1998) and in the `true' IGM 
($N_{\HI}<10^{14}$ cm$^{-2}$) at $z\sim 5$ (Songaila 2002).
Identification of C IV, Si IV and O VI absorption lines which correspond to 
Ly$\alpha$ absorption lines in the spectra of high redshift quasars has 
revealed that the low density IGM has been enriched up to 
$Z_{IGM}\sim 10^{-2.5\pm0.5}\; Z_{\odot}$. For lower column density clouds (the
`true' IGM) Songaila (2002) found a lower limit on the IGM metallicity at
$z\sim 5$ of $10^{-3.5}\; Z_{\odot}$.

With the assumed star formation efficiency ($f_\star=0.534$ for a Salpeter; 
$f_{\star}=0.214$ for a heavy IMF; $f_{\star}=0.142$ for a very heavy IMF)
the IGM would be enriched to the observed
metallicity already at $z= 15-18$ for the Ly$\alpha$ forest value and
at $z =20-25$ for the `true' IGM lower limit (Fig. \ref{fig:metal}). 
As seen from Fig. \ref{fig:metal}, for $z\sim8.8$ 
the mean IGM metallicity exceeds by one order of magnitude that  
observed in the Ly$\alpha$ forest.

\begin{figure}
\center{{
\epsfig{figure=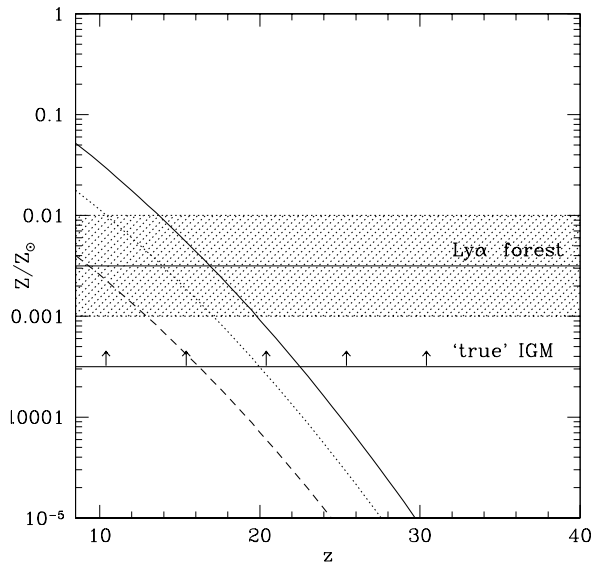,height=8cm}
}}
\caption{\label{fig:metal_yy} As Fig. \ref{fig:metal} but in the case in 
which only pair-instability SNe eject metals into the IGM. 
{\it Dashed line}: Salpeter IMF. 
{\it Dotted line}: heavy IMF. {\it Solid line}: very heavy IMF.}
\end{figure}

As an alternative, we are thus forced to consider the analogous case in which only 
pair-instability SNe eject metals into the IGM (Fig. \ref{fig:metal_yy}). 
For a Salpeter IMF the obtained metallicity is now consistent with the limits 
of the Ly$\alpha$ forest, whereas for a heavy IMF this value is slightly exceeded. 
For top heavy IMFs, for which the contribution of normal SNe is not so 
important, the results change only slightly.  

In order to further reduce these metallicity values, we have to consider lower 
star formation efficiencies. Consequently only a fraction of the
NIRB can be due to Pop III stars. By imposing that Pop III stars do not
eject more metals than the observed ones, we find self consistent limits on the
star formation efficiency and on the  
contribution of the first population of stars to the NIRB; Table
\ref{tab:frac} summarizes the results. Apart for the case of a Salpeter IMF in which only 
pair-instability SNe contribute to the metal enrichment of the IGM, the 
star formation efficiency has to be small ($\simlt 10$\%) and the fraction of 
NIRB due to Pop III stars is $\simlt 20$\%. 

The value obtained by Songaila (2002) for the `true' IGM at $z\sim5$ 
allows us to set a lower limit
on the star formation efficiency and the contribution of Pop III stars to the 
NIRB. The efficiency has to be greater than few$\times10^{-4}$ to enrich the 
IGM to $Z\sim10^{-3.5}\; Z_{\odot}$; thus, the Pop III star contribution to 
the NIRB is at least of the order of 0.4-1.7\%. This fraction increases up to
8\% if we consider the case in which only pair-instability SNe contribute
to the metal enrichment of the universe.

\begin{table}
\begin{center}
\begin{tabular}{llcc}
\hline
\hline
IMF & & $f_\star$ & \% of NIRB \\

\hline
Salpeter & {\it all SN}  & $9.1\times10^{-2}$ & 17\% \\
 & {\it only SN$_{\gamma\gamma}$} & $4.3\times10^{-1}$ & 80\%  \\
Heavy & {\it all SN}  & $1.5\times10^{-2}$ & 7\% \\
 & {\it only SN$_{\gamma\gamma}$} & $3.8\times10^{-2}$ & 18\%  \\
Very Heavy & {\it all SN}  & $0.6\times10^{-2}$ & 4\%  \\
 & {\it only SN$_{\gamma\gamma}$} & $0.8\times10^{-2}$ & 6\%   \\
\hline
\end{tabular}
\caption{The star formation efficiency and the fraction of 
NIRB from Pop III stars obtained by imposing 
that metals ejected by Pop III SNe do not exceed the mean observed
values in the Ly$\alpha$ forest. The {\bf cut} 
model for the galaxy counts is assumed.}
\label{tab:frac}
\end{center}
\end{table}

\section{Discussion}

Pop III stars can account for the entire NIRB excess        
if the high redshift star formation efficiency is $f_\star=10\%-50\%$, 
depending on the IMF and on the unknown `normal' galaxy contribution 
in the L band. 
Our best results are obtained by considering a value $z_{end}\approx 8.8$ for the
redshift at which the formation of Pop III stars ends, almost independently
of the IMF and of the `normal' galaxy contribution model. A hard upper limit 
$z_{end} \le 9$ is set by the J band data.  
The above efficiency values are not at odd with expectations
for objects forming in the Dark Ages (Madau, Ferrara \& Rees 2001), 
although they seem to be higher than those found at lower redshifts.

In addition, the values of $f_\star$ found should be 
considered as upper limits since the unaccounted NIRB level could be lower
due to a different zodiacal light subtraction. To quantify this uncertainty
we have performed fits using the Wright zodiacal light model. 
In this case the DIRBE data alone are considered,
as NIRS data are only available for the Kelsall model.
We find that required star formation efficiencies decrease to $0.15-0.41$ 
($z_{end}=8-10$) with a Salpeter IMF, $0.06-0.16$ with a heavy IMF, and 
$0.04-0.1$  with a heavy IMF;
at $z_{end}=8.8$ the resulting star formation efficiencies are 0.23 
(Salpeter IMF), 0.09  (heavy IMF), and 0.06 (very heavy IMF).
However, the errors are doubled with respect to the case of Kelsall model for the 
zodiacal light emission.
Also, the $\chi^2$ analysis does not allow to evaluate a unique 
redshift at which Pop III formation ends. This is due to
the fact that the break around 1 $\mu$m is not as sharp as in the
previous case, as already mentioned. 


As a consistency test, we have calculated the associated IGM metal
enrichment.  If the whole
unaccounted NIRB is due to Pop III stars, then we predict that the 
low density (`true') IGM would be enriched at the
observed mean value already  by $z=15-20$, and that the Ly$\alpha$ 
forest data imply that either a Salpeter or a heavy IMF with characteristic mass
$M_c =15 M_\odot$ are acceptable, provided pair-instability SNe are the only
source of heavy elements. This conclusion can be modified by the two following
occurrences: {\it (i)} only stars with masses in excess of 260 $M_\odot$ are
formed, that lock    their nucleosynthetic products into the VMBH remnant,
effectively breaking the light-metal production link (Schneider et al. 2002) or 
{\it (ii)} metals are 
very inhomogeneously mixed in the IGM, with filling factors much smaller than unity
(Scannapieco, Ferrara \& Madau 2002) at these high redshifts.
In both cases, the limits obtained from metallicity arguments are much 
weaker if not purely indicative. 

To exemplify case {\it (i)}, let us consider 
an IMF of the form $\phi(M)=\delta(1000\; \Msun)$. By imposing that the
whole NIRB is due to Pop III stars ($f_\star=0.04\pm0.01$ and 
$z_{end}=8.83\pm0.01$) and assuming the {\bf cut}
model for the galaxy counts, we obtain a density of VMBH
$\Omega_{VMBH}(z=8.8)\sim7\times10^{-4}$.
This value is two orders of magnitude greater than
the measured mass density in super massive black holes (SMBHs) found from
the demography of nuclei of nearby galaxies (Magorrian et al. 1998; Gebhardt et
al. 2001), $\Omega_{SMBH}=10^{-4}h\Omega_b=2.66\times10^{-6}$ 
(Merritt \& Ferrarese 2001), but it is unlikely that all  VMBHs merge into
SMBHs (Schneider et al. 2002; Volonteri, Haardt \& Madau 2002). Hence it is not clear 
if this scenario can represent a viable solution.  

In the second case {\it (ii)}, the measured metallicity might not be representative of
the actual cosmic metal density and might result in a large overestimate
of the amount of IGM heavy elements. In addition,  
note that above estimates assume that all the produced heavy elements
can escape into the IGM. If the metal escape fraction is equal or less than  
$\sim 10$\%, the whole unaccounted NIRB could easily be due to Pop III stars
with no conflict with the observed IGM metallicity limits. 

As a final remark, we point out that 
additional sources might contribute to the NIRB, as accretion onto
black holes (Carr 1994) or decay of massive primordial particles (Bond,
Carr \& Hogan 1986).





\section*{Acknowledgements}
We thank D.~Schaerer for providing Pop III spectra and  V. Bromm and L.~Danese for 
constructive and stimulating comments. 
This work was partially supported (AF) by the Research and Training Network
`The Physics of the Intergalactic Medium' set up by the European Community
under the contract HPRN-CT2000-00126 RG29185.


\end{document}